\documentclass{elsart}

\usepackage{amssymb}
\usepackage[dvips]{epsfig}
\usepackage{color}

\begin{document}

\begin{frontmatter}

\title{Induced charge-density oscillations at metal surfaces}

\author[DIPC]{V. M. Silkin\corauthref{cor_Silkin}},
\ead{waxslavs@sc.ehu.es}
\author[DIPC]{I. A. Nechaev},
\author[DIPC,MIXTO]{E. V. Chulkov}, and
\author[DIPC,MIXTO]{P. M. Echenique}
\corauth[cor_Silkin]{Corresponding author. Tel.: +34 943 018284; Fax.: +34 943 015600}
\address[DIPC]{Donostia International Physics Center (DIPC), P. de Manuel Lardizabal 4,
20018 San Sebasti{\'a}n, Basque Country, Spain}
\address[MIXTO]{Departamento de F\'{\i}sica de Materiales, Facultad de
Ciencias Qu\'{\i}micas, UPV/EHU and Centro Mixto CSIC-UPV/EHU, Apdo. 1072, 20080 San Sebasti{\'a}n, Basque
Country, Spain}

\begin{abstract}

Induced charge-density (ICD) oscillations at the Cu(111) surface caused by an external impurity are studied
within linear response theory. The calculation takes into account such properties of the Cu(111) surface
electronic structure as an energy gap for three-dimensional (3D) bulk electrons and a $s-p_z$ surface state
that forms two-dimensional (2D) electron system. It is demonstrated that the coexistence of these 2D and 3D
electron systems has profound impact on the ICD in the surface region. In the case of a static impurity the
characteristic ICD oscillations with the $1/\rho^2$ decay as a function of lateral distance, $\rho$, are
established in both electron systems. For the impurity with a periodically time-varying potential, the novel
dominant ICD oscillations which fall off like $\sim1/\rho$ are predicted.

\end{abstract}

\begin{keyword}
Surface electronic phenomena \sep Noble metals \sep Low index single metal surfaces \sep Electronic surface
states \sep Adsorbates 
\end{keyword}

\end{frontmatter}

In the last years, scanning tunneling microscopy has become a powerful tool to study phenomena governed by
screening at metal surfaces, i.e., charge rearrangement in response to the disturbance caused by impurities
or defects
\cite{crlus93,haavprl93,sppes97,hobrprl97,waekapa98,sataprb99,boovprl00,malun00,niwaprl03,remeprl04}.
Remarkable examples of that are the visualization of surface-state-originated Friedel oscillations of the
induced charge-density (ICD) at metal surfaces \cite{crlus93,haavprl93} and the investigation of the indirect
long-range interaction between adsorbates mediated by these oscillations \cite{ei96}. It has been
demonstrated \cite{remoprl00,knbrprb02} that at noble-metal (111) surfaces this interaction, whose energy
decays as $1/\rho^2$ with the lateral distance $\rho$ between adsorbates, can lead to mutual redistribution
of adsorbed atoms up to $\rho$ $\sim10$ nm. This interaction might also be responsible for the rearrangement
of adsorbate species on the Si(111)-$\sqrt{3}\times\sqrt{3}$-Ag surface \cite{sanaprb99}. Moreover, very
recently it has been reported on a realization of a self-assembled two-dimensional (2D) atomic structure due
to this interaction \cite{sopiprl04}.

In the theoretical studies of such phenomena, the scattering approach has found the wide application
\cite{remoprl00,hypejp00,hecrn94,fiheprl01}. Within this approach, a $s-p_z$ surface state at the noble-metal
(111) surfaces is considered to form a 2D free electron gas, ignoring the fine structure of the surface-state
wave function and the fact that this 2D electron gas ${\it coexists}$ with underlying three-dimensional (3D)
electron system. It is well known that the 2D electron system responds to the introduction of an impurity by
producing Friedel oscillations with the characteristic $1/\rho^2$ decay \cite{lakoss78}, whereas in the 3D
electron gas the oscillations fall off as $1/R^3$, $R$ being a distance from the impurity \cite{lavoj60}.
Nevertheless, up to date, the question of how the 2D and 3D systems respond ${\it simultaneously}$ in the
vicinity of a metal surface has not been addressed yet. An answer can be obtained, in principle, from
\textit{ab initio} calculations. But, at present, these evaluations are feasible only for systems with the
lateral distances between adsorbates of the order of several {\AA} \cite{fiscprl00}.

In this Letter we investigate the response of an electron system at a metal surface to the introduction of an
impurity, considering  the Cu(111) surface as an example. We show that taking explicitly into account the
realistic Cu(111) surface band structure, which is characterized by the energy gap at the center of surface
Brillouin zone (SBZ) and the $s-p_z$ partly occupied surface state, is crucial for the description of the
surface response.

In order to evaluate the ICD, we adopt an approach based on linear response theory in which an external
perturbation $V_{ext}({\bf r}';\omega)$ and the corresponding ICD $n_{ind}({\bf r};\omega)$ are related by
the equation
\begin{equation}
n_{ind}({\bf r};\omega) =  \int d {\bf r}'  \chi({\bf r},{\bf r}';\omega) V_{ext}({\bf r}';\omega).
\label{nindr}
\end{equation}
Here $\chi({\bf r},{\bf r}';\omega)$ is the density-response function which is non-local and complex and
contains information on electronic excitations. In time-dependent density functional theory $\chi({\bf
r},{\bf r}';\omega)$ satisfies the integral equation \cite{pegoprl96}
\begin{equation}\label{dyson}
\chi = \chi^0 +  \chi^0(v_c+K_{xc})\chi
\end{equation}
with $\chi^0({\bf r},{\bf r}';\omega)$ being the response function of a noninteracting electron system,
${v_c}({\bf r}-{\bf r}')$ is the Coulomb  potential, and $K_{xc}({\bf r},{\bf r}';\omega)$ accounts for
dynamical exchange-correlation effects. As we are mainly interested in the evaluation of long-range charge
density oscillations, we use the random phase approximation, i.e., $K_{xc}=0$. The inclusion of a non-zero
$K_{xc}$ mainly affects the amplitude of Friedel oscillations \cite{egprb85,khteprb02,gisissc03}.

\begin{figure}[tbp]
\centering
\includegraphics[scale=0.4,angle=0]{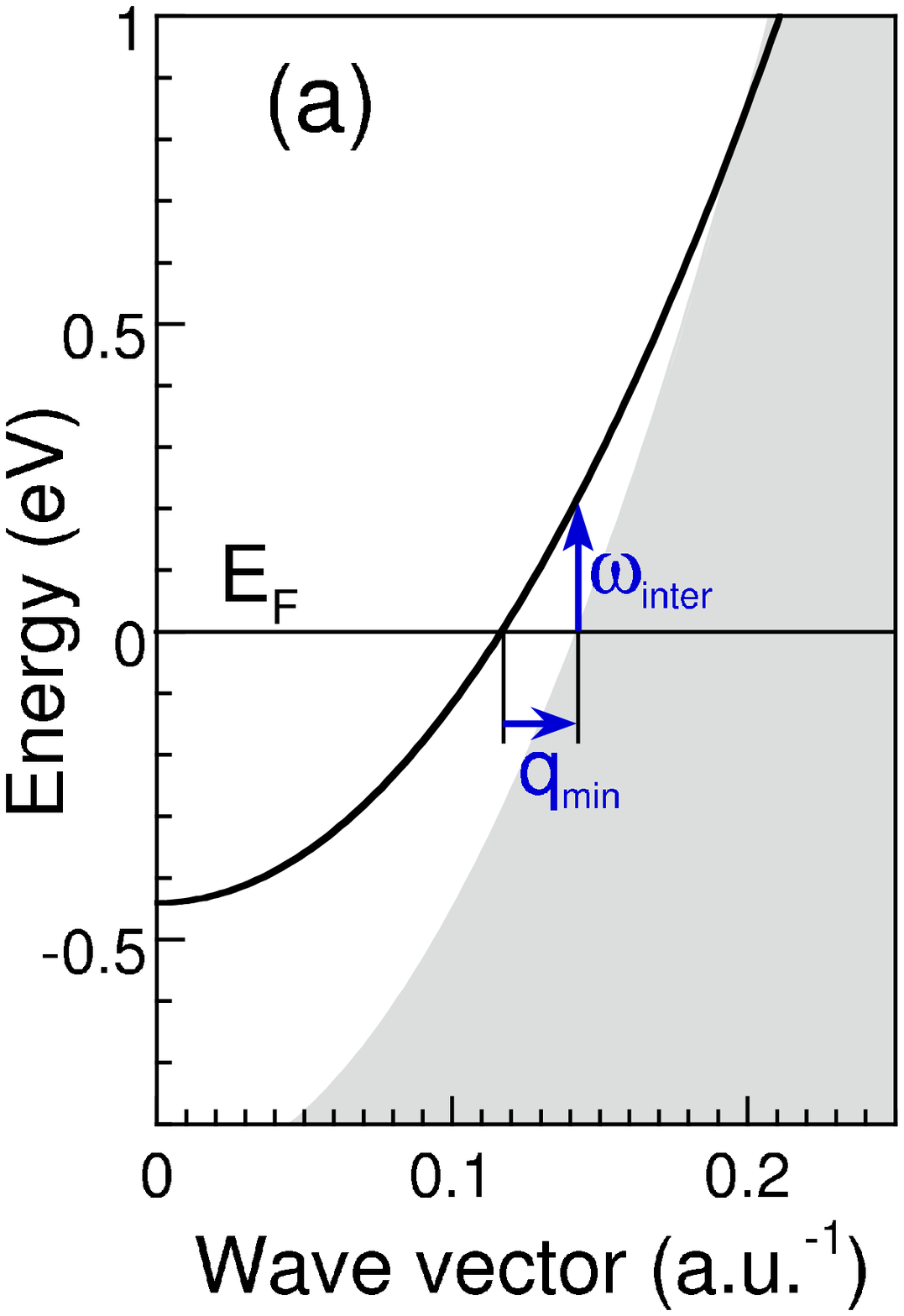}
\includegraphics[scale=0.4,angle=0]{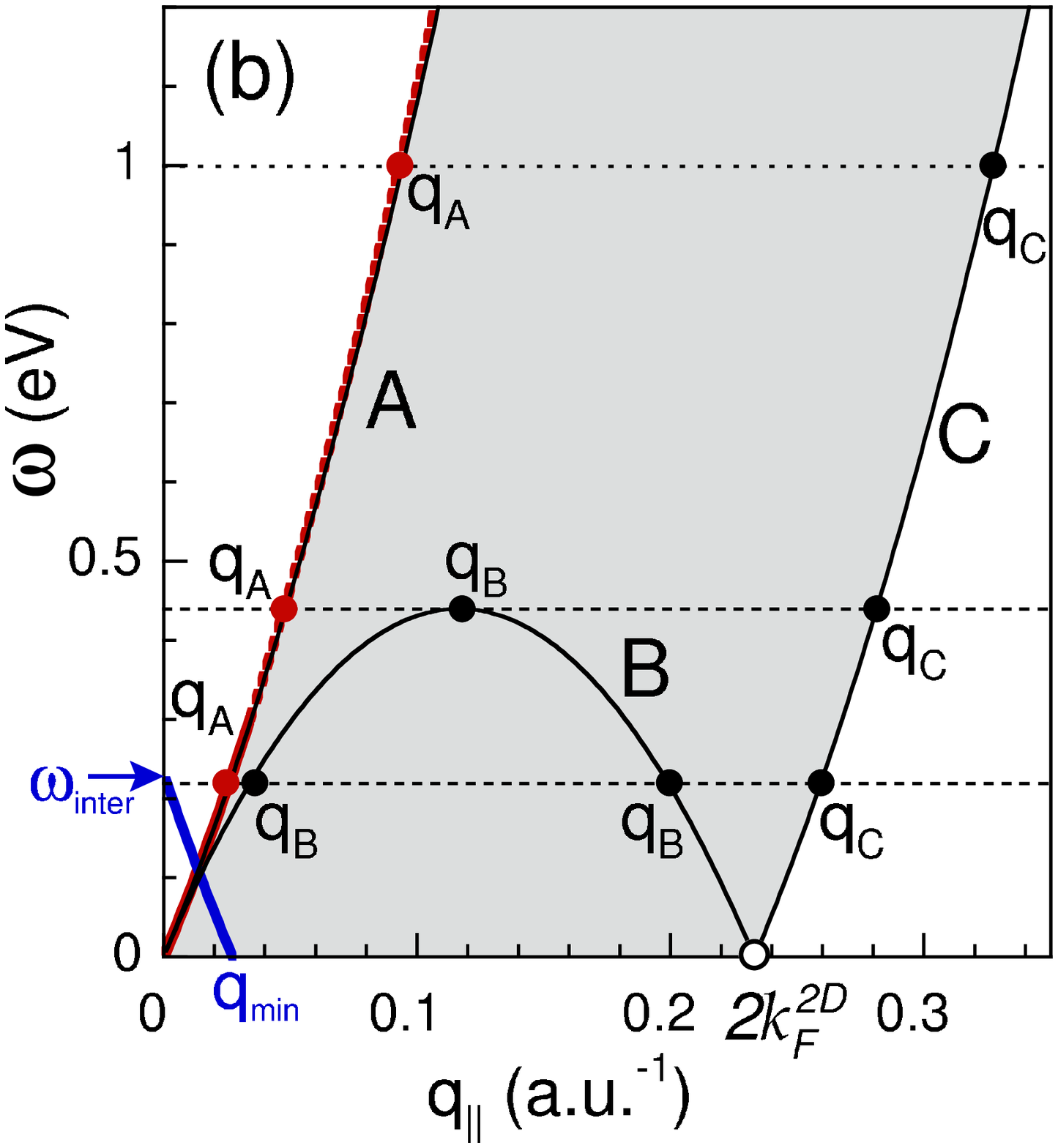}\\
\caption{ (a) Surface electronic structure of Cu(111). The grey area represents the projected bulk electronic
structure. Solid line shows the surface state dispersion with an effective mass of $m^*=0.42m_e$. (b) The
phase space available for $e-h$ excitations. In the grey area the surface state intraband $e-h$ excitations
are permitted. The intraband $e-h$ bulk excitations are possible everywhere. The interband ``surface
state---bulk'' transitions are forbidden in the area below the blue line. Lines A, B, and C are respectively
described by equations $\omega=v_F^{2D}q_{||}+q_{||}^2/2m^{*}$, $\omega=v_F^{2D}q_{||}-q_{||}^2/2m^{*}$, and
$\omega=-v_F^{2D}q_{||}+q_{||}^2/2m^{*}$ with $v_F^{2D}$ being the 2D (surface state) Fermi velocity. Red
line shows the acoustic surface plasmon (ASP) dispersion \cite{sigaepl04} very close to the line A (solid red
line corresponds to the well defined ASP, whereas dashed one indicates the region of its gradual
degradation).} \label{states_phase}
\end{figure}

To describe the Cu(111) surface, we employ a slab containing 81 atomic layers of Cu together with a vacuum
region corresponding to 20 interlayer spacings. The substrate is described by a model one-dimensional
potential of Ref. \cite{chsiss97}. This potential reproduces the key ingredients of the Cu(111) electronic
structure shown in Fig.~\ref{states_phase}(a), namely, the energy gap at the center of the SBZ and the
$s-p_z$ surface and image-potential states in it. Using translation invariance parallel to the surface, one
can perform the 2D Fourier transform for all quantities in Eq. (\ref{nindr}). The ICD has now the following
form \cite{egprb85}
\begin{eqnarray}
n_{ind}(\rho,z;\omega) =  \frac{1}{2\pi}  \int_{0}^{\infty} d q_{||} q_{||}
J_0(q_{||}\rho)n_{ind}(q_{||},z;\omega), \label{nind} \\
n_{ind}(q_{||},z;\omega)=\int dz'\chi(q_{||},z,z';\omega)V_{ext}(q_{||},z';\omega),\label{nind_f}
\end{eqnarray}
where $V_{ext}(q_{||},z';\omega)$ is the 2D Fourier transform of an external potential, $J_0$ is the Bessel
function of order $0$, and $z$ ($\rho$) is a distance perpendicular (parallel) to the surface. $\chi^0$ is
given by\footnote{Hartree atomic units are used throughout unless otherwise is stated.} 
\begin{eqnarray}
\chi^{0}(q_{||},z,z^{\prime };\omega )&=&2\mathrel{\mathop{\sum }\limits_{i,j}}\phi _{i}(z)\phi _{j}^{\ast
}(z)\phi _{j}(z^{\prime })\phi
_{i}^{\ast }(z^{\prime })  \nonumber \\
&\times& \int \frac{d{\bf k}_{||}}{(2\pi )^{2}}%
\frac{f_{i,{\bf k}_{||}+{\bf q}_{||}}-f_{j,{\bf k}_{||}}} {E_{i,{\bf k}_{||}+{\bf q}_{||}} -E_{j,{\bf
k}_{||}}+\omega +i\eta },  \label{chi0}
\end{eqnarray}
where the sum runs over the bands $i$ and $j$, $f$'s are the Fermi factors, $E_{i,{\bf k}_{||}+{\bf
q}_{||}}=\varepsilon_i+({\bf k}_{||}+{\bf q}_{||})^2/2m_i^*$, $E_{j,{\bf k}_{||}}=\varepsilon_j+{\bf
k}_{||}^2/2m_j^*$. $\varepsilon_{i}$ ($\phi_{i}$) represent the one-particle energies (wave functions)
obtained with the model potential \cite{chsiss97}. The experimental values of effective masses $m_{i}^*$
different from the free electron mass have been used in Eq. (\ref{chi0}). The regions of possible
electron-hole ($e-h$) excitations at Cu(111) are shown in Fig.~\ref{states_phase}(b). In the calculation of
$\chi^0$, a finite value for the broadening parameter $\eta=0.001$ eV was introduced.

\begin{figure}[tbp]
\centering
\includegraphics[scale=0.55,angle=270]{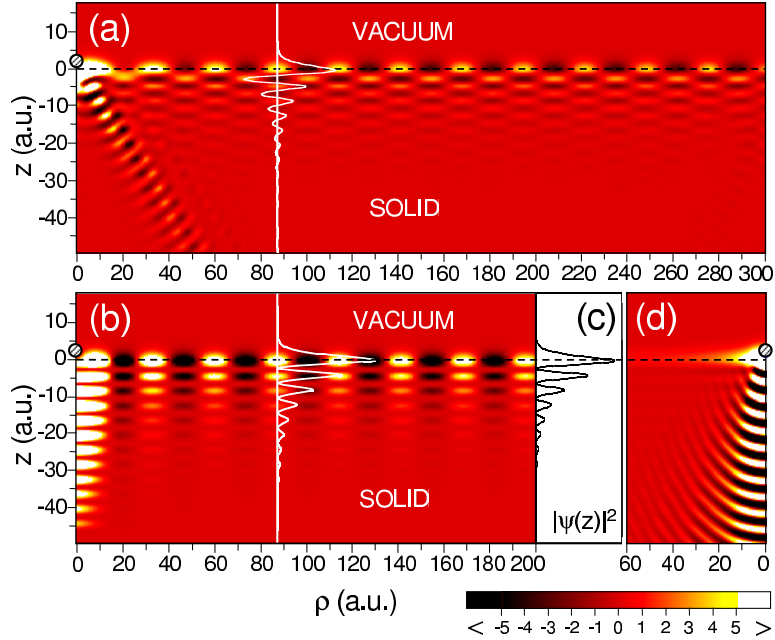}\\
\caption{ (a) Map of $n_{ind}\times(\rho^2+(z-z_0)^2)$ (in a.u.$\times10^5$) induced by an impurity placed at
the point shown by shaded circle as a function of $z$ and $\rho$. Horizontal dashed line shows the crystal
border. The solid (vacuum) is located for $z<0$ ($z>0$). The surface atomic layer is at $z=-1.97$ a.u. (b)
The same as in (a) for a hypothetical case of a free standing surface state without bulk electrons.
Additionally, in (a, b) $n_{ind}\times(\rho^2+(z-z_0)^2)$ as a function of $z$ is shown by white curves for
the vertical cut $\rho=87$ a.u. (c) The surface state charge density as a function of $z$. (d) Shows the same
as in (a) for the jellium model. Note the absence of lateral ICD oscillations in this case. }
\label{static_map}
\end{figure}

We begin with the case of response to a static external potential of the form
\begin{equation}
V_{ext}({\bf r}) = - Qe / |{\bf r}-{\bf r}_0|, \label{vext}
\end{equation}
caused by an impurity of charge $Qe=1$ placed at ${\bf r}_0=\{{\bf r}_{||}=0,z_0\}$.\footnote{In all figures, the impurity was placed at $z_0=1.97$ a.u.} 
Fig.~\ref{static_map}(a) shows the obtained ICD multiplied by $\rho^2+(z-z_0)^2$ as a function of $z$ and
$\rho$. One can see that in a horizontal plane the Friedel oscillations peculiar to 2D electron systems are
nicely reproduced: the ICD amplitude shows the quadratic dependence on $\rho$ in the lateral direction and
the period of the oscillations is precisely determined by the Fermi wave vector of the surface state,
$k_F^{2D}=0.1165$ a.u. In the direction perpendicular to the surface the ICD oscillations demonstrate the
behavior more complicated than that expected for a 3D electron gas.
Note also that any displacement of the impurity along the  $z$-direction in the vicinity of the crystal
border changes only the amplitude of the oscillations but not their character.

To reveal the importance of the 3D system for the ICD shown in Fig.~\ref{static_map}(a), we consider a
hypothetical case of a free standing surface state without bulk electrons (the case of a pure 2D electron
gas). Fig.~\ref{static_map}(b) shows the resulting ICD. As in Fig.~\ref{static_map}(a), one can observe the
2D Friedel oscillations in the lateral direction, while in the perpendicular direction the ICD behavior is in
clear correspondence with the location of surface-state charge-density presented in Fig.~\ref{static_map}(c).
However, there is a qualitative difference between the oscillations presented in Figs.~\ref{static_map}(a)
and ~\ref{static_map}(b): the ICD as a function of $z$ in Fig.~\ref{static_map}(a) changes sign whereas in
the pure 2D case it does not occur (see white curves in Figs.~\ref{static_map}(a) and (b) which give the ICD
along the vertical cut $\rho=87$ a.u.). Similar behavior is observed for any $\rho$. Inspecting
Fig.~\ref{static_map}(a), one can see that the $\sim1/\rho^2$ decay takes place both in the 2D electron
system and in the region of the 3D system adjacent to the crystal border. In the latter system for a given
$z$ the ICD oscillates with the opposite sign to that in the 2D one. To clarify the origin of this ICD
behavior we separate contributions to $\chi_0$ in Eq. (\ref{chi0}) into three parts: "surface state---surface
state" ($s-s$), "bulk---bulk" ($b-b$), and "surface state---bulk" ($s-b$) transitions
\begin{equation}\label{chi_zero_sum}
\chi^{0}(q_{||},z,z^{\prime };\omega ) = \sum_{\alpha} \chi^{0}_{\alpha}(q_{||},z,z^{\prime };\omega ),
\end{equation}
where $\alpha$ stands for $s-s$, $b-b$, or $s-b$, and the summation is performing over these three
components. As a result, the ICD of Eq.~(\ref{nind_f}) can be rewritten as the sum of the partial ICD
concerned with the contributions of the 2D and 3D systems as well as their mixing
\begin{equation}\nonumber
n_{ind}(q_{||},z;\omega ) =\sum_{\alpha} n_{ind}^{\alpha}(q_{||},z;\omega ),
\end{equation}
where
\begin{eqnarray}
n_{ind}^{\alpha}(q_{||},z;\omega ) &=&
  \int d z'
\chi^{\alpha}(q_{||},z,z';\omega)\{V_{ext}(q_{||},z';\omega) \nonumber\\
&+& \int d z''v_c(q_{||},z',z'')\sum_{\beta\ne\alpha} n_{ind}^{\beta}(q_{||},z'';\omega)\}. \label{n_all}
\end{eqnarray}
Here $v_c(q_{||},z',z'')=\frac{2\pi}{q_{||}}e^{-q_{||}|z'-z''|}$ is the 2D Fourier transform of the bare
Coulomb interaction, each $\chi^{\alpha}$ is defined through Eq.~(\ref{dyson}) by the corresponding
$\chi_{\alpha}^0$.

Thus, the ICD shown in Fig.~\ref{static_map}(a) comprises three interrelated components $n_{ind}^{\alpha}$
which represent the response of the corresponding ``subsystems'' to both the external perturbation
(\ref{vext}) and the electrostatic field created by the two other $n_{ind}^{\beta}$. Note that if we neglect
the $\chi^{0}_{s-b}$ term in Eq.~(\ref{chi_zero_sum}) and approximate the surface-state wave function by the
$\delta$-function, we obtain the results of Ref.~\cite{pinaprb04}. The main advantage of Eq.~(\ref{n_all})
over Eq.~(\ref{nind_f}) is the possibility to solve it iteratively and to analyze step-by-step how one
component affects the rest and vice versa. Actually, by setting $n_{ind}^{b-b}=n_{ind}^{s-b}=0$, we find
$n_{ind}^{s-s}$ shown in Fig. \ref{static_map}(b). It can be considered as a first iteration. Then, by
substituting the obtained $n_{ind}^{s-s}$ into $n_{ind}^{b-b}$ with $n_{ind}^{s-b}=0$, we find the response
of the 3D system to the external field and the field created by the ICD of Fig. \ref{static_map}(b), and so
on. It explains why the $\sim1/\rho^2$ decay of the ICD peculiar to the 2D system takes also place in the 3D
electron system adjacent to the crystal border. Additionally the screening by the 3D system of the complex
perturbation mentioned above is the origin of an alternate behavior of the ICD as a function of $z$ at a
given $\rho$ (see the white line in Fig.~\ref{static_map}(a)).

To emphasize the role of the realistic electronic structure in the surface response, we plot in
Fig.~\ref{static_map}(d) the ICD obtained within a jellium model for $r_s=2.67$ commonly used for description
of Cu $sp$-valence electrons. In this case the distinct behavior of $n_{ind}$ along the crystal surface
without any oscillations is clearly seen. Instead, this model gives strong ICD oscillations in the direction
perpendicular to the surface which are suppressed in the realistic consideration (Fig.~\ref{static_map}(a))
due to complicated mutual influence of one ``subsystem'' on the others.\footnote{More detailed analysis will be given elsewhere.} 
Also, at small $\rho$, this influence leads to the ICD penetrating into the bulk only along a straight line
at a finite angle away from the surface normal. Similar effect has also been observed in the studies of
long-lived adsorbate-induced states at metal surfaces \cite{gabofd00,bokaprb02}.
\begin{figure}[tbp]
\centering
\includegraphics[scale=0.55,angle=0]{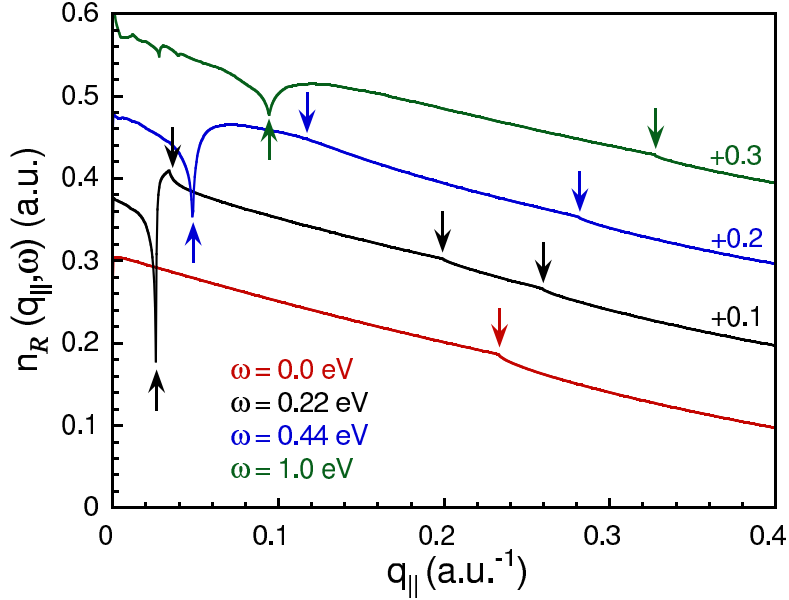}\\
\caption{  $n_{R}(q_{||},z=0;\omega_0)$ vs $q_{||}$ for four values of $\omega_0$. $n_R$ for $\omega_0=0.22$
eV, $0.44$ eV, and $1.0$ eV are displaced vertically by $0.1$, $0.2$, and $0.3$ a.u., respectively. The
arrows $\uparrow(\downarrow)$ indicate the singularities positions at $q_A(q_{B,C})$. } \label{ro_ind_mom}
\end{figure}

Now we consider the case of the potential (\ref{vext}) varying in time with some small frequency $\omega_0$
as $\cos(\omega_0 t)$. As we deal with the linear system, the superposition principle would hold true.
Therefore, once the response to the given potential is known, the response to an arbitrary time-varying
external potential can be evaluated.  In contrast to the previous static case when the integration in Eq.
(\ref{nind}) over $q_{||}$ is performed along the line $\omega=0$ of Fig.~\ref{states_phase}(b), here the
integration should be performed along the line $\omega=\omega_0$. Moreover, in this case the ICD depends on
time and has the following form
\begin{eqnarray} \label{dynamic_ind_den}
n_{ind}(\rho,z;t)&=&n_R(\rho,z;\omega_0)\cos(\omega_0t)+n_I(\rho,z;\omega_0)\sin(\omega_0t) \nonumber\\
&=&A(\rho,z;\omega_0)\cos(\omega_0t-\theta(\rho,z;\omega_0)) \nonumber
\end{eqnarray}
where $n_R(\rho,z;\omega_0)$ and $n_I(\rho,z;\omega_0)$ are defined through Eqs.~(\ref{nind}) and
(\ref{nind_f}) by the real and imaginary parts of the response function $\chi(q_{||},z,z';\omega_0)$,
respectively. The ICD amplitude is $A=\sqrt{n_R^2+n_I^2}$ and the phase shift $\theta$ is defined by
$\tan(\theta)=n_I/n_R$. Thus, a non-vanishing $n_I$ leads to a non-zero phase shift between the ICD and
external perturbation oscillations (it means an energy absorption in the electron system).

Fig.~\ref{states_phase}(b) shows, as an example, three lines corresponding to $\omega_0=0.22$ eV, $0.44$ eV,
and $1.0$ eV. Contrary to the ``static'' $\omega_0=0$ line which has only one singularity point at
$q_{||}=2k_F^{2D}$ ensuring the Friedel oscillations in $n_{ind}$ with the $1/\rho^2$ decay, the
$\omega_0\neq 0$ lines have up to four singularity points (labeled by $q_{A,B,C}$) due to the singularities
at lines A, B, and C \cite{somehandbook}, what makes the dynamic response more complex.
\begin{figure}[tbp]
\centering
\includegraphics[scale=0.55,angle=270]{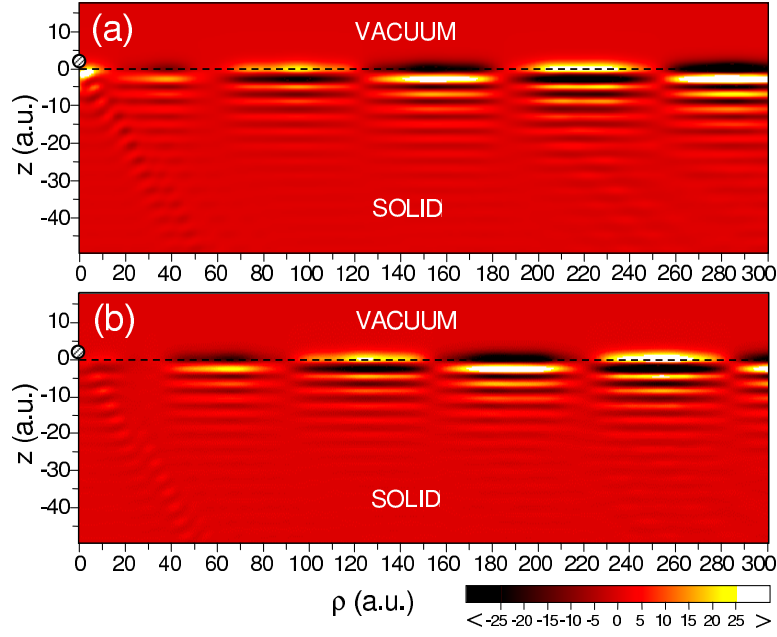}\\
\caption{ Map  of $n_R$ (a) and $n_I$ (b)
  induced by the time-varying external
perturbation (\ref{vext}) with frequency $\omega_0=0.44$ eV
  as a function of $z$ and
$\rho$ and multiplied by $\rho^2+(z-z_0)^2$. Note the enhanced scale comparing with the one of Fig.
\ref{static_map}. Other notions are the same as in Fig. \ref{static_map}(a). } \label{dynamic_044}
\end{figure}
\begin{figure}[tbp]
\centering
\includegraphics[scale=0.55,angle=0]{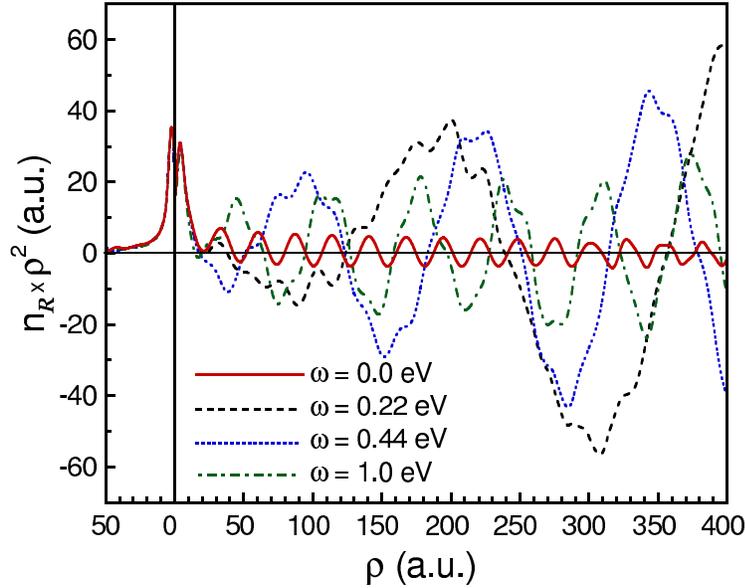}\\
\caption{ $n_{R}(\rho,z=0;\omega_0)$ multiplied by $\rho^2$ vs $\rho$ for $\omega_0=0.0$ eV, $0.22$ eV,
$0.44$ eV, and $1.0$ eV. On the left of the vertical line $\rho=0$, the corresponding curves in the jellium
model do not show any oscillations. } \label{ro_ind_dist}
\end{figure}
Commonly, these singularities are similar to that at $q_{||}=2k_F^{2D}$ in the static case. But, at the
Cu(111) surface the  $q_A$ singularity corresponds to the location of the acoustic surface plasmon (ASP) (see
Fig.~\ref{states_phase}(b)), whose origin and nature has recently been studied \cite{sigaepl04,pinaprb04}.
The ASP singularity, appearing due to the coexistence of 2D and 3D electron systems \cite{pinaprb04}, is much
stronger than the other ones and leads to the $\sim\cos(q_A\rho)/\rho$ -like oscillations, i.e., with the
spatial decay inversely proportional to $\rho$ only. To visualize the distinct nature of singularities at
$q_A$ and $q_{B,C}$, the dependence of $n_{R}$ on $q_{||}$ for $z=0$ is shown in Fig.~\ref{ro_ind_mom}.

Fig.~\ref{dynamic_044} represents the ICD  components $n_R$ (ICD at the moment $t=2n\pi/\omega_0$, $n$ is
integer) and $n_I$ ($t=(2n+1/2)\pi/\omega_0$) for $\omega_0=0.44$ eV. One can clearly see that in this case
the ICD along the surface is dominated by oscillations with $q_A$ wave vector corresponding to the ASP.
Nevertheless, the ICD also contains contributions originated from the other two singularity points, $q_B$ and
$q_C$. In general, the structure of the resulting oscillations is quite complicated, but fitted rather well
by  three $\cos$-like functions. To compare the decay law of oscillations, we show $n_{R}(\rho,z=0;\omega_0)$
for $\omega_0=0.0$ eV, $0.22$ eV, $0.44$ eV, and $1.0$ eV in Fig.~\ref{ro_ind_dist}.\footnote{Some distortion
seen in the $\omega_0=0$ ICD for $\rho \geq 300$ a.u. is explained by finite size of the slab. An increase of
the slab thickness moves gradually this disturbance region farther
away in $\rho$.} 
For small $\omega_0$, the oscillations in 2D with $q_A=\omega_0/v_F^{2D}$ have the $\sim \cos(q_A\rho)/\rho$
asymptotic behavior. The same kind of ICD oscillations arises in the 3D system  in the proximity of the 2D
electron one.   As $\omega_0$ increases, the ASP looses its strength due to interband transitions (see
Figs.~\ref{states_phase}(b) and ~\ref{ro_ind_mom}), and oscillations with $q_A$ gradually change their decay
behavior from $\rho^{-1}$ to $\rho^{-2}$. Thus, for $\omega_0=0.22$ eV the decay obeys the $\rho^{-1}$ law,
while it is proportional to $\rho^{-1.4}$ for $\omega_0=0.44$ eV, and $\rho^{-1.7}$ for $\omega_0=1.0$ eV.

In conclusion, we nicely reproduce within linear response theory the Friedel oscillations of the ICD at the
Cu(111) surface  caused by a static impurity, taking explicitly into account the realistic surface band
structure. It is shown that the coexistence of the 2D electron system with the 3D one has profound impact on
the screening properties at the metal surface. Thus, additionally to the $\sim\cos(k_F^{2D}\rho)/\rho^2$ ICD
oscillations in the 2D system along the surface, the same kind of oscillations occurs in the 3D electron
system in the region adjacent to the crystal border.  In the case of a time-varying potential, dominant
$\cos(\omega_0 / v_F^{2D} \cdot \rho)/\rho$ -like oscillations  at the surface in both the 2D and 3D electron
systems are predicted for low frequencies. We expect that these oscillations can lead to a longer-range
indirect interaction between atoms and molecules at metal surfaces than it is thought nowadays.

\section*{Acknowledgments}

The authors thank A. Arnau  and I. Nagy  for useful discussions. We acknowledge partial support by the
University of the Basque Country, the Departamento de Educaci\'on del Gobierno Vasco, MCyT (Grant No. MAT
2001-0946) and the European FP6 Network of Excellence (FP6-NoE NANOQUANTA (500198-2)).


\begin{thebibliography}{9}

\bibitem{crlus93}  M.F. Crommie, C.P. Lutz, and D.M. Eigler, Science {\bf 262} (1993) 218.

\bibitem{haavprl93}  Y. Hasegawa and P. Avouris, Phys. Rev. Lett. {\bf 71} (1993) 1071.

\bibitem{sppes97} P.T. Sprunger, L. Petersen,  E.W. Plummer,  E. L{\ae}gsgaard,
and F. Besenbacher, Science {\bf 275} (1997) 1764.

\bibitem{hobrprl97}  Ph. Hofmann,  B.G. Briner, M. Doering, H.-P. Rust, E.W. Plummer,  and A.M. Bradshaw,
Phys. Rev. Lett. {\bf 79} (1997) 265.

\bibitem{waekapa98} E. Wahlstr\"om, I. Ekwall, H. Olin, and L. Walld\'en,  Appl. Phys. A {\bf 66} (1998) S1107.

\bibitem{sataprb99} N. Sato, S. Takeda, T. Nagao, and S. Hasegawa,  Phys. Rev. B {\bf 59} (1999) 2035.

\bibitem{boovprl00} A. Begicevic, S. Ovesson, P. Hyldgaard, B.I. Lundqvist, H. Brune, and D.R. Jennison,
Phys. Rev. Lett. {\bf 85} (2000) 1910.

\bibitem{malun00} H.C. Manoharan, C.P. Lutz, and D.M. Eigler, Nature (London) {\bf 403} (2000) 512.

\bibitem{niwaprl03} N. Nilius, T.M. Wallis, M. Persson, and W. Ho, Phys. Rev. Lett. {\bf 90} (2003) 196103.

\bibitem{remeprl04} J. Repp, G. Meyer, and K.-H. Rieder, Phys. Rev. Lett. {\bf 92} (2004) 036803.

\bibitem{ei96} T.L. Einstein, Handbook of Surface Science, ed. W.N. Unertl (Elsevier,
Amsterdam, 1996), Vol. 1, p. 577.

\bibitem{remoprl00} J. Repp, F. Moresco, G. Meyer, K.-H. Rieder, P. Hyldgaard, and M. Persson,
  Phys. Rev. Lett. {\bf 85} (2000) 2981.

\bibitem{knbrprb02} H. Knorr, H. Brune, M. Epple, A. Hirstein, M.A. Schneider, and K. Kern,
  Phys. Rev. B {\bf 65} (2002) 115420.

\bibitem{sanaprb99} N. Sato, T. Nagao, and S.Hasegawa, Phys. Rev. B {\bf 60} (1999) 16083.

\bibitem{sopiprl04} F. Silly, M. Pivetta, M. Ternes, F. Patthey, J.P. Pelz, and W.-D. Schneider,
  Phys. Rev. Lett. {\bf 92} (2004) 016101.

\bibitem{hypejp00} P. Hyldgaard and M. Persson, J. Phys.: Condens. Matter {\bf 12} (2000) L13.

\bibitem{hecrn94} E.J. Heller, M.F. Crommie, C.P. Lutz, and D.M. Eigler, Nature (London) {\bf 369} (1994) 464.

\bibitem{fiheprl01} G.A. Fiete, J.S. Hersch, E.J. Heller, H.C. Manoharan, C.P. Lutz, and D.M. Eigler, Phys. Rev. Lett.
{\bf 86} (2001) 2392.

\bibitem{lakoss78} K.W. Lau and W. Kohn, Surf. Sci. {\bf 75} (1978) 691.

\bibitem{lavoj60} J.S. Langer and S.H. Vosko, J. Phys. Chem. of Solids {\bf 12} (1960) 196.

\bibitem{fiscprl00} K.A. Fichthorn and M. Scheffler, Phys. Rev. Lett. {\bf 84} (2000) 5371.

\bibitem{pegoprl96} M. Petersilka, U.J. Gossmann, and E.K.U. Gross, Phys. Rev. Lett. {\bf 76} (1996) 1212.

\bibitem{egprb85}  A. G. Eguiluz, Phys. Rev. B {\bf 31} (1985) 3303.

\bibitem{khteprb02} I.G. Khalil, M. Teter, and N.W. Ashcroft , Phys. Rev. B {\bf 65} (2002) 195309.

\bibitem{gisissc03} G.F. Giuliani and G.E. Simion, Solid State Commun. {\bf 127} (2003) 789.

\bibitem{chsiss97} E.V. Chulkov,  V.M. Silkin, and P.M. Echenique, Surf. Sci. {\bf 391} (1997) L1217;
Surf. Sci. {\bf 437} (1999) 330.

\bibitem{sigaepl04} V.M. Silkin, A. Garc\'{\i}a-Lekue, J.M. Pitarke, E.V. Chulkov, E. Zaremba, and P.M. Echenique,
Europhys. Lett. {\bf 66} (2004) 260.



\bibitem{pinaprb04} J.M. Pitarke, V.U. Nazarov, V.M. Silkin, E.V. Chulkov, E. Zaremba, and P.M. Echenique,
Phys. Rev. B {\bf 70} (2004) 205403.


\bibitem{gabofd00} J. P. Gauyacq, A. G. Borisov, G. Raseev, and A. K. Kazansky, Faraday Discuss. {\bf 117} (2000) 15.

\bibitem{bokaprb02} A. G. Borisov, A. K. Kazansky, and J. P. Gauyacq,  Phys. Rev. B {\bf 65} (2002) 205414.

\bibitem{somehandbook} M.C. Desjonqu\`{e}res and D. Spanjaard, Concepts in Surface Physics,
Springer-Verlag, Berlin, 1996.


\end{thebibliography}
\end{document}